# Computer Simulation Study of the Levy Flight Process

Mehrdad Ghaemi [1,*], Zahra Zabihinpour [1,†], Yazdan Asgari [2,‡]

[1] *Department of chemistry, Tarbiat Moallem University, Tehran, Iran*

[2] *Center for Complex Systems Research, K.N.Toosi University of Technology, Tehran, Iran*

**Abstract.** Random walk simulation of the Levy flight shows a linear relation between the mean square displacement $<r^2>$ and time. We have analyzed different aspects of this linearity. It is shown that the restriction of jump length to a maximum value ($l_m$) affects the diffusion coefficient, even though it remains constant for $l_m$ greater than 1464. So, this factor has no effect on the linearity. In addition, it is shown that the number of samples does not affect the results. We have demonstrated that the relation between the mean square displacement and time remains linear in a continuous space, while continuous variables just reduce the diffusion coefficient. The results are also implied that the movement of a levy flight particle is similar to the case the particle moves in each time step with an average length of jumping $<l>$. Finally, it is shown that the non-linear relation of the Levy flight will be satisfied if we use time average instead of ensemble average. The difference between time average and ensemble average results points that the Levy distribution may be a non-ergodic distribution.

*PACS codes and Keywords*: 05.40.Fb, 82.20.Wt, Levy flight, Anomalous diffusion, Nonlinearity, Non-ergodicity.

## 1. Introduction

It has been shown that the anomalous diffusion plays a fundamental role in the dynamics of a wide class of systems. So it has recently been the topic of extensive re-searches [1]. Turbulent flow [2], phase-space motion in chaotic dynamics [3] and transport in highly heterogeneous media such as porous materials or gels [1,4] are some of the main instances where anomalous diffusion underlies transport processes. The Anomalous diffusion is characterized by a non-linear growth of the mean square displacement (ensemble average over trajectories), i.e.

[*] Corresponding author, Tel: (+98)2188848949 e-mail: ghaemi@tmu.ac.ir
[†] e-mail: zabihin@gmail.com
[‡] e-mail: yazdan1130@gmail.com

$$\langle r(t)^2 \rangle \propto t^\alpha \tag{1}$$

for $\alpha \neq 1$, relation(1) deviates from the linear dependence $<r(t)^2> \propto t$ that characterizes a normal diffusion. Interested readers can find further information about theoretical background of the relation (1) in a good review article by Metzler and Klafter [5].

Two somewhat different approaches can be used to model the anomalous diffusion by means of random walks. One approach, in the framework of the continuous time formulation of random walk, makes use of long tailed waiting time distributions [4,6].The other approach considers long tailed distributions for jump length probability and discrete time steps[7]. Discrete time random walks with long tailed jump distributions are paradigmatically represented by the Levy flights [4-8].

The use of Levy flights in understanding of nature is a quite broad practice reaching fields as apparently different as physics and economy or being useful in the explanation of phenomena happening at the biological level at several levels of complexity. In particular the work regarding the diffusion of micelles [9] is not only a place where Levy flight was used to model but the first experimental finding of a Levy flight. Levy flight was used not only to model but also to explain and understand the broad presence of large fluctuations in several and quite different areas of knowledge such as econophysics [10] or human neuroscience [11].

The Levy flight is defined by a jump probability $p(r)$ whose Fourier transform reads $p(\mathbf{k})=\exp(-ak^\gamma)$, with $k=|\mathbf{k}|$ and $\gamma<2$. For $\gamma=2$, a normal diffusion recovered. Since it is not possible to find a closed form for the corresponding $p(r)$, it is usually replaced by a function with the same asymptotic behavior. A possible form for the jump probability is [12]

$$p(r) \propto \frac{1}{r^{d+\gamma}} \tag{2}$$

where $r=|r|$ and $d$ is the spatial dimension. The Levy Flight represents a special class of the discrete Markovian processes for which the mean square displacement diverges. In order to make the mean square displacement a useful quantity for the analysis, some restrictions have to be imposed on the spatial temporal stepping distributions. For finite times, i.e. in a certain portion of the whole random walk, and on averages over a finite number of trajectories the mean square displacement for a Levy flyer is determined as follow [12]

$$\langle r^2 \rangle \propto t^{\frac{2}{\gamma}} \tag{3}$$

In 1996, Tribel and Boon [13] have used the Lattice Gas Automata (LGA) for the simulation of the Levy flight. They have found the relation between the mean square displacement and time is linear. They have also mentioned the linearity is due to the discreteness of time and space. In this article, simulation is carried out for a single particle which jumps according to the Levy distribution. Therefore there is no difference between the lattice gas automata and ordinary random walk methods. We have analyzed different aspects which lead to the linearity of the relation between the mean square displacement and time. The obtained linear relation may be arisen from the following restrictions: a) using small maximum jump length ($l_m$) instead of large jumps in the relation (2), b) insufficient number of samples, c) using integers for $r$ instead of continuous numbers, and d) the method of averaging. In section 2, we have shown the restriction of jump length, limited number of samples, and the method of computation i.e. using integers for $r$ or using continuous numbers based on the Weron algorithm, cannot affect the linear relation between the mean square displacement and time. But the method of time averaging can produce the non-linear relation between the mean square displacement and time. That is to say, the non-linearity of the Levy flight will be satisfied if we use a time average instead of an ensemble average.

## 2. Numerical Simulation of the Levy Flight

A two-dimensional discrete time random walk whose underlying jump length distribution exhibits the form of the relation (2) with $\gamma = 0.5$ is considered where $\gamma$ is the Levy exponent. Although there is an exact analytical representation for $\gamma=0.5$ [14], we face with restrictions of dimensions and discrete time in computations. To end up the computations, we need to truncate the space domain. So, the jump distribution has a cutoff for sufficiently large $r$ ($r \in [1, l_m]$). According to the relation (3) and the value of the Levy exponent ($\gamma=0.5$), it is expected

$$<r^2> \propto t^{\frac{2}{\gamma}} = t^4 \tag{4}$$

A particle was located at the center of a square lattice with $10^6 \times 10^6$ cells and could jump to the right, left, up, or down direction with the length $r$. The length of jumping $r$ was an integer variable which satisfied the Levy distribution function (relation 2) and the minimum and maximum length jump was set equal to 1 (lower cutoff) and 1500 ($l_m$), respectively. The process was done for $10^4$ time steps and the simulation was repeated for $2 \times 10^3$ samples. The Mersenne Twister pseudorandom number generator with period 2^19937-1 was used for generating uniform random numbers. The result is shown in Figure 1. As it is clearly shown, there is a linear relation between the mean square displacement and time with the diffusion constant equal to 107.23 for this case. In other words, an anomalous diffusion was not observed, which is in agreement with the result obtained by Tribel and Boon [13].

The obtained linear relation may be arisen from the following restrictions: a) using small $l_m$ instead of large jumps in the relation (2), b) insufficient number of samples, c) using integers for $r$ instead of continuous numbers, and d) the method of averaging (the ensemble average was used in the previous simulation).

For considering the effect of the maximum length jump ($l_m$), the simulation was repeated for different $l_m$. The results displayed in Figures 2 and 3 show that the relation between the mean square displacement and time remains linear, even though the diffusion coefficient increases with increasing the value of $l_m$. The results in Figure 2 also show that the diffusion coefficient remains constant for $l_m$ greater than 1464.

In the next step, the simulation was repeated for different numbers of samples (500, 1000, 2000, and 3000 samples). However, the linearity still remained and the calculated diffusion coefficient was 107.23 for all cases.

We have removed the limitation of the integer jumps by using the Weron algorithm [15] for generating random variables according to the Levy distribution function in a continuous space. So, a particle was located at the center of a two dimensional Cartesian coordinate. In each time step, the particle jumped to a location which was on the circumference of a circle with radius $r$, where $r$ was a continuous Levy variable and $l_m$=1500. The particle moved $10^4$ steps and the simulation was done for $2\times10^3$ samples. As it is demonstrated in Figure 4, in spite of using Levy distribution in a continuous space, the anomalous diffusion is not observed. However, the diffusion constant is reduced to 24.68.

In the next step, the value of the average length of jumping $<l>$ for $2\times10^3$ samples and $10^4$ time steps with $l_m$=1500 was calculated. The calculated value for the average length of jumping $<l>$ was 14.59. After that, a simulation for a normal diffusion with the length of jumping equal to $<l>$ =14.59 was done. The results are shown in Figure 5. As it is seen, the method of the ensemble averaging (with $l_m$=1500) for the Levy flight movement of a particle is similar to the case in which the particle moves with an average length of jumping $<l>$ in each time step.

Therefore, just one more factor remains to be considered: the method of the averaging. We had used the ensemble average in the previous calculations. So, the simulations should be repeated using the time average in one sample for both continuous and discrete spaces.

First, the simulation was considered in a $10^6\times10^6$ square lattice with a fixed boundary condition. A particle was located at the center of the lattice. It moved to one of the up, down, left or right direction (with an equal probability) with the length of $r$, where $r$ was considered as an integer Levy distributed variable which was truncated to $l_m$=1500. The particle moved $13\times10^3$ time steps and $r^2$ was averaged each $10^3$ steps. The results in Figure (6-a) show a non-linear relation between the mean square displacement and time, even though they do not satisfy the relation (4) correctly. Next, the Weron algorithm was used for generating random variables according to the Levy distribution function in a continuous space. A particle was located at the center of a two-dimensional Cartesian coordinate. In each time step, the particle jumped to a location which was on the circumference of a circle with radius $r$, where $r$ was a continuous Levy variable and $l_m$=1500. It moved $13\times10^3$ time steps and $r^2$ was averaged each $10^3$ steps. That is to say, values of $r^2$ are added for each 1000 time steps and the result is divided by 1000 in order to obtain time average for each point. The results are shown in Figure (6-b) which demonstrate a non-linear relation between the mean square displacement and time. After that, we have fitted the results according to a power equation $y=ax^b$ with $a = 4.3 \times10^{-8}$ and $b = 3.91$ (see the curve in Figure (6-b)). So, the result correctly satisfies the relation (4) for $b\approx 2/\gamma$.

For considering the long time behavior of this approach, we repeat the simulation in a continuous space based on the Weron algorithm up to $4\times10^6$ time steps and $r^2$ was averaged similar to the previous case but in each $10^5$ time steps. As it is seen in Figure 7, the results are still compatible with the relation (4) for $b \approx 2/\gamma$.

Therefore, in order to obtain the expected value for γ in relation (4), one may consider a system with continuous space, even though the time averaging could lead to the non-linearity of the system in both cases.

## 3. Conclusion

Our results show that the restriction of jump length, limited number of samples, and the method of computation, i.e. using integers for *r* (an ordinary random walk algorithm) or using continuous numbers (the Weron algorithm) cannot affect the linear relation between the mean square displacement and time. It is shown that the linearity could be removed using time average instead of ensemble average.

It seems there are other time average methods which may lead to a similar result. But we do not know clearly which methods will lead to a precise result. However, there are some problems such as effect of step sizes on average calculation, using coarse graining method, and number of time steps which could affect simulation results. At this time we cannot provide a deeper explanation about the use of other methods for averaging, because there are some fundamental problems that must be answered before anything. For example: for finite time step different random seeds can produce different results. So, how a single trajectory must be selected from infinite possible trajectories? We are pursuing in order to answer these questions in our further works.

The difference between time average and ensemble average result points that the Levy distribution may be a non-ergodic distribution. We are not the first who point to the possibility of non-ergodicity of Levy flight, since it has been considered as a problem by Zumofen et. al [16] and others[17,18] in recent years.

## Acknowledgment

The authors acknowledge Prof. R. Islampour for his useful comments.

**Figure Captions:**

**Figure 1:** The mean square displacement versus time for the Levy flight simulation (a square lattice with $10^6 \times 10^6$ cells, $l_m$=1500, $10^4$ time steps, $2 \times 10^3$ samples)

**Figure 2:** The diffusion coefficient for different lengths of jumping (The result are obtained from a square lattice with $10^6 \times 10^6$ cells, $10^4$ time steps, and $2 \times 10^3$ samples)

**Figure 3:** The mean square displacement versus time for the Levy flight simulation for different $l_m$: a) $l_m$=500, b) $l_m$=1000, and c) $l_m$=1500. ($10^4$ time steps, $2 \times 10^3$ samples)

**Figure 4:** The mean square displacement versus time using the Weron algorithm for generating random variables in a continuous space ($10^4$ time steps, $2 \times 10^3$ samples)

**Figure 5:** The mean square displacement versus time for two cases: *a*) the Levy flight simulation, $D_a$=107.23 ($l_m$=1500, $10^4$ time steps, $2 \times 10^3$ samples). *b*) The particle moves like a normal diffusion with the length of jumping equal to $L=<l>$=14.59, $D_b$=106.71 ($10^4$ time steps, $2 \times 10^3$ samples)

**Figure 6:** The mean square displacement versus time using the time averaging: a) time averaging based on integer Levy distributed variable. ($13 \times 10^3$ time steps, $r^2$ is averaged each $10^3$ steps). b) time averaging based on the Weron algorithm ($13 \times 10^3$ time steps, $r^2$ is averaged each $10^3$ steps). The results are fitted to a power equation $y=ax^b$ where $a$=4.3$\times 10^{-8}$, $b$=3.91

**Figure 7:** The mean square displacement versus time based on the Weron algorithm in the case of long time behavior ($4 \times 10^6$ time steps, $r^2$ is averaged each $10^5$ steps). The results are fitted to a power equation $y=ax^b$ where $a$=3.37$\times 10^{-13}$, $b$=4.0

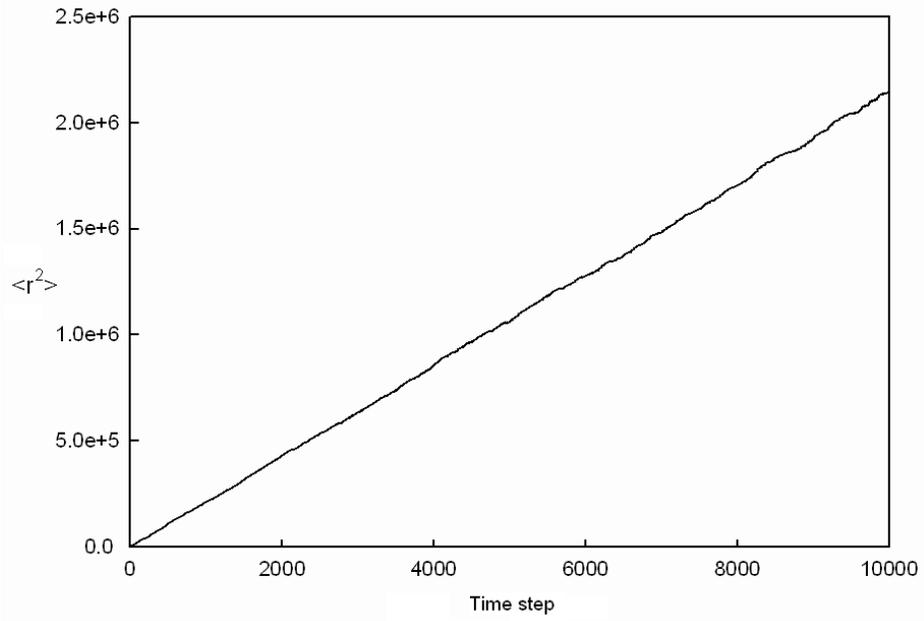

Figure 1

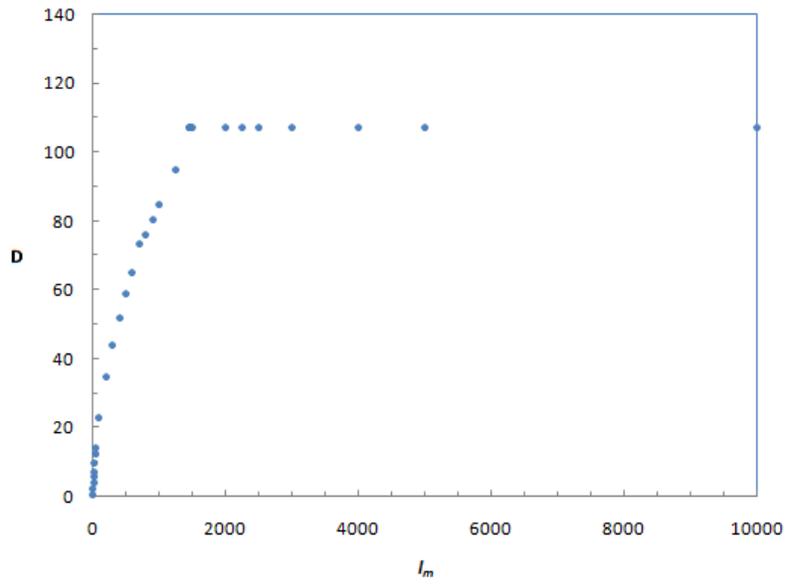

Figure 2

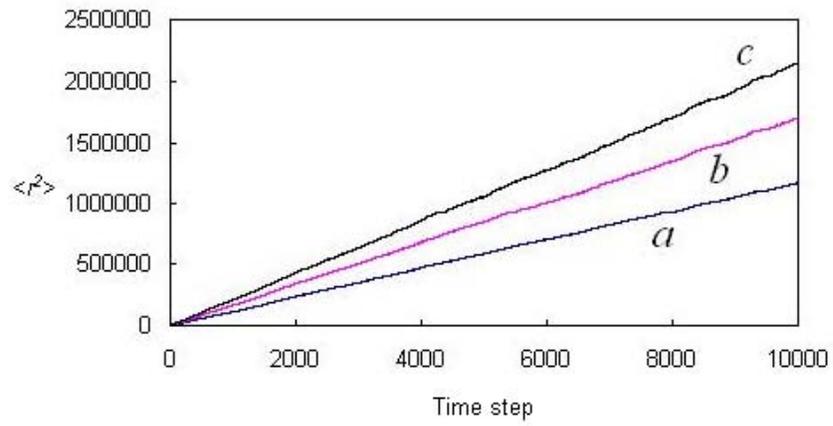

Figure3

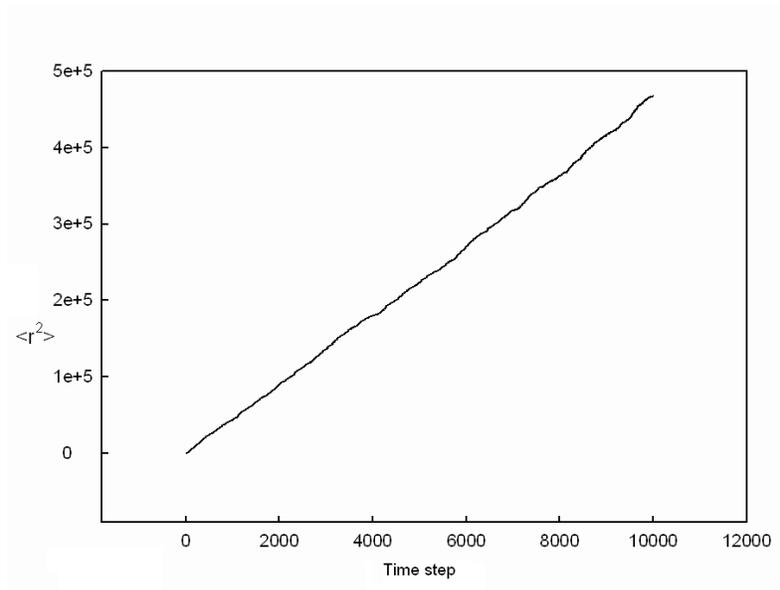

Figure 4

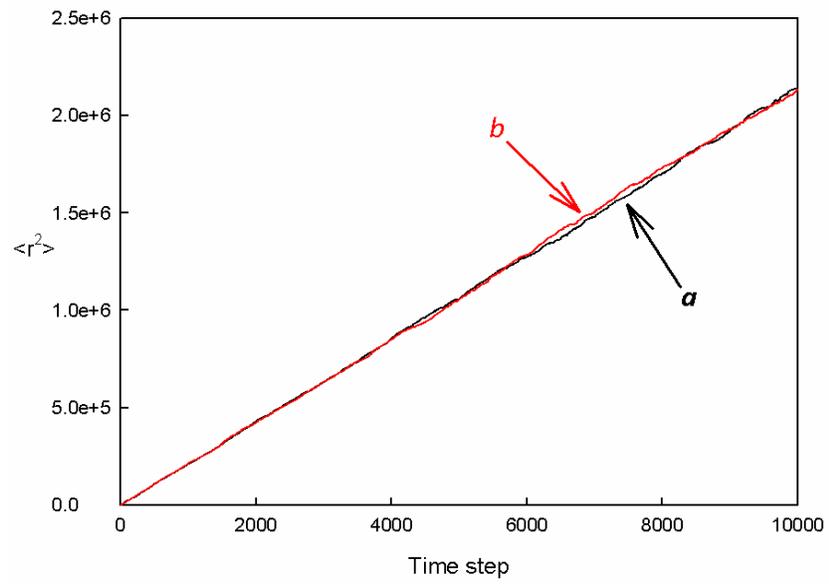

Figure 5

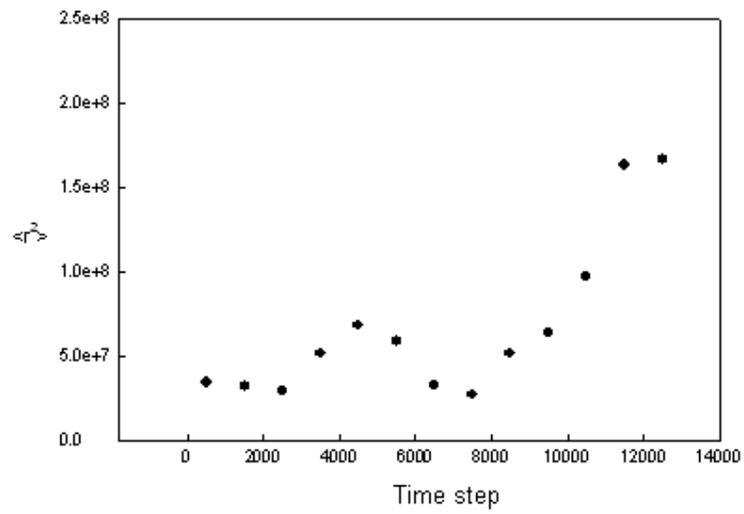

Figure 6a

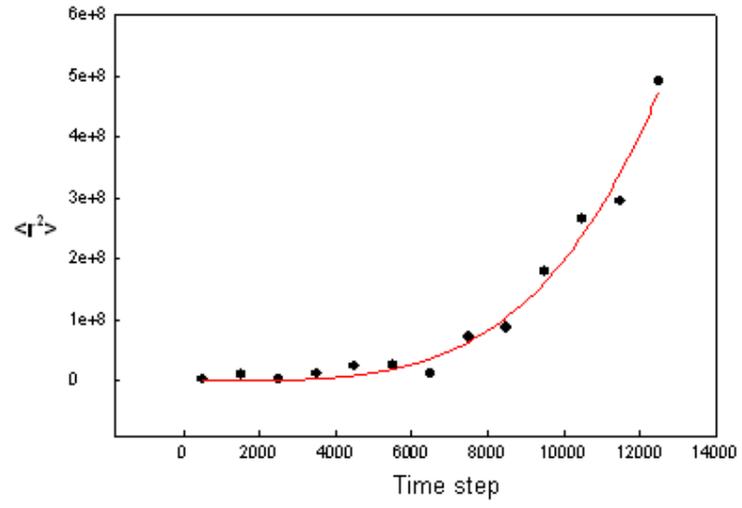

Figure 6b

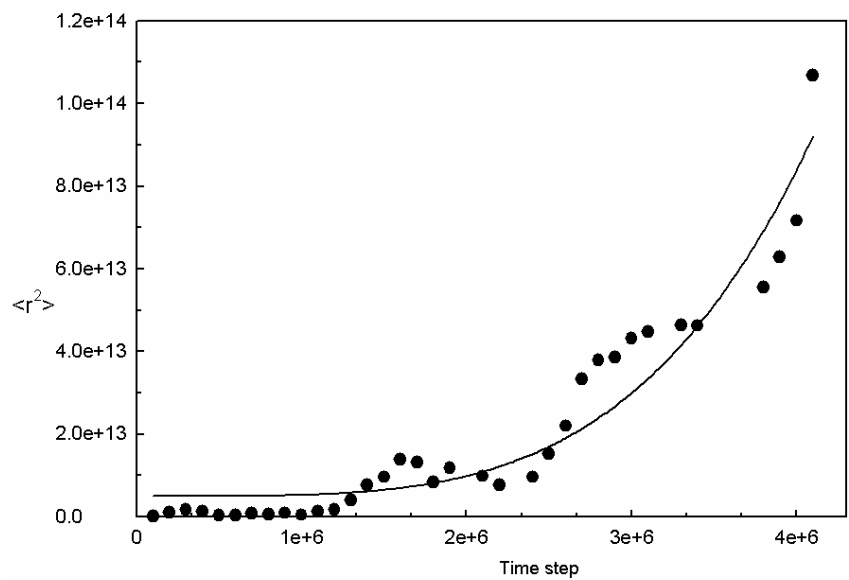

Figure 7